\documentclass[conference]{IEEEtran}
\usepackage{amsmath,amssymb,amsthm}
\usepackage{graphics, subfigure}
\usepackage{tikz}
\usepackage{pgfplots}
\usepackage{tabularx}
\usepackage{algorithm, algorithmic}
\usepackage{color}
\usepackage{multirow}
\usetikzlibrary{automata,positioning}
\usepackage{wrapfig}
\usepackage{rotating}
\usepackage{flushend}
\usepackage{bbm}

\usepackage{algorithm, algorithmic}

\usepackage{epstopdf, xspace}

\usepackage{mathtools}

\usepackage{pgfplots} 
\pgfplotsset{compat=newest} 
\pgfplotsset{plot coordinates/math parser=false} 
\newlength\figureheight 
\newlength\figurewidth

\begin{document}
\definecolor{brown}{cmyk}{0,0.81,1,0.60}
\definecolor{magenta}{rgb}{0.4,0.7,0}
\definecolor{gray}{rgb}{0.5,0.5,0.5}
\definecolor{red}{rgb}{1,0,0}
\definecolor{green}{rgb}{0.5,0,0.5}
\definecolor{blue}{rgb}{0,0,1}


\ifthenelse{\isundefined{\final}} {
\newcommand{\vaneet}[1]{{\color{green}(VA: #1)}}
\newcommand{\shuai}[1]{{\color{red}(SH: #1)}}
\newcommand{\feng}[1]{{\color{blue}(FQ: #1)}}
\newcommand{\anis}[1]{{\color{brown}(AE: #1)}}
\newcommand{\shubho}[1]{{\color{magenta}(SS: #1)}}
}{
\newcommand{\vaneet}[1]{}
\newcommand{\shuai}[1]{}
\newcommand{\feng}[1]{}
\newcommand{\anis}[1]{}
\newcommand{\shubho}[1]{}
}

\newcommand{\eat}[1]{}

\newcommand{\BULLET}{\vspace{+.03in} \noindent $\bullet$ \hspace{+.00in}}

\title{Reinforcement Learning Based Scheduling Algorithm for 
Optimizing Age of Information in Ultra Reliable Low Latency Networks }

 \author{Anis Elgabli, Hamza Khan, Mounssif Krouka, and Mehdi Bennis\\ Center of Wireless Communications\\University of Oulu, Oulu, Finland} 

\maketitle


\if
$a_n(k+1)=a_n(k)+\sum_{i=1}^{d_n^{k}}\Big(\sum_{t=NACK_{i-1}}^{T_{i}}1_{L_{n,i}^t}\Big)+ACK$

$L_{n,i}^t=L_{n,i-1}^t-t \cdot r(t)$
\vspace{0.1in}

$L_{n,i}^{NACK_{i-1}}=L_{n}$
\vspace{0.1in}

$\textbf{Maximize: } \frac{1}{KN}\sum_{n=1}^N\sum_{k=1}^K \big(a_n(k)-a_n(k-1)\big)+\cdots$
\vspace{0.1in}

subject to
\vspace{0.1in}

$a_n(k) > a_n(k-1), \forall n,k$
\fi
\vspace{1cm}
\begin{abstract}
Age of Information (AoI) measures the freshness of the information at a remote location. AoI reflects the time that is elapsed since the generation of the packet by a transmitter. In this paper, we consider a remote monitoring problem (e.g., remote factory) in which a number of sensor nodes are transmitting time sensitive measurements to a remote monitoring site. We consider minimizing a metric that maintains a trade-off between minimizing the sum of the expected AoI of all sensors and minimizing an Ultra Reliable Low Latency Communication (URLLC) term. The URLLC term is considered to ensure that the probability the AoI of each sensor exceeds a predefined threshold is minimized. Moreover, we assume that sensors tolerate different threshold values and generate packets at different sizes. Motivated by the success of machine learning in solving large networking problems at low complexity, we develop a low complexity reinforcement learning based algorithm to solve the proposed formulation. We trained our algorithm using the state-of-the-art actor-critic algorithm over a set of public bandwidth traces. Simulation results show that the proposed algorithm outperforms the considered baselines in terms of minimizing the expected AoI and the threshold violation of each sensor.
\end{abstract}
\begin{IEEEkeywords}
AoI, URLLC, Stochastic Optimization, Reinforcement Learning
\end{IEEEkeywords}
\section{Introduction}
\label{sec:intro}
Cyber-Physical Systems (CPS), Internet of Things (IoT), Remote Automation, and Haptic Internet are all examples for systems and applications that require real-time monitoring and low-latency constrained information delivery. The growth of time sensitive information led to a new data freshness measure named {\it Age of Information} (AoI). AoI measures packet freshness at the destination accounting for the time elapsed since the last update generated by a given source \cite{realtimeKaul}.

Consider a cyber-physical system such as an automated factory where a number of sensors are located at a remote site and they are transmitting time sensitive information to a remote observer through the internet (multihop wired and wireless network). Each sensor's job is to sample measurements from a physical phenomena and transmit them to the monitoring site. Due to the variability of the network bandwidth because of many reasons such as the resource competition, packet corruptions, and wireless channel variations, maintaining fresh data at the monitoring side become challenging problem.

Recently, there were few papers tackling the problem of minimizing the AoI of number of sources that are competing for the available resources. \cite{OptimAoI} considers the problem of many sensors connected wirelessly to a single monitoring node and formulates an optimization problem that minimizes the weighted expected AoI of the sensors at the monitoring node. Moreover, the authors of \cite{packetdeadline} also consider the sum expected AoI minimization problem when constraints on the packet deadlines are imposed. In \cite{AoIShared}, sum expected AoI minimization is considered in a cognitive shared access.

However, the problem of AoI minimization for ultra reliable low latency communication (URLLC) systems should pay more attention to the tail behavior in addition to optimizing on average metrics~\cite{khayri2018v2v}. In URLLC, the probability that the AoI of each node sharing the resources exceeds a certain predetermined threshold should be minimized. Therefore, the AoI minimization metric for URLLC should account for the trade-off between minimizing the expected AoI of each node and maintaining the probability that the AoI of each node exceeding a predefined threshold at its minimum. Therefore, in this paper, we consider optimizing a metric that is a weighted sum of two metrics (i) Expected AoI, and (ii) probability of threshold violation. 

Introducing the second metric in the object function, accounting for the fact that different nodes are generating packets at different sizes, and different nodes can tolerate different threshold violation percentages. Moreover, accounting for the non-causality knowledge of the available bandwidth, and including the non zero packet drop that might be encountered in the network. All these factors, contribute to the complexity of the problem. In other words, the proposed scheduling problem is a stochastic optimization problem with integer non convex constraint which is in general hard to solve in polynomial time. 

Therefore, motivated by the success of machine learning in solving many of the online large scale networking problem when trained offline. In this paper, we propose an algorithm based on Reinforcement Learning (RL) to solve the proposed formulation. RL is defined by three components (state, action, and reward). Given a state, the RL agent is trained in offline manner to choose the action that maximizes the system reward. The RL agent interacts with the environment in a continuous way and tries to find the best policy based on the reward/cost fed back from that environment \cite{ML2018}. In other words, the RL agent tries to choose the trajectory of actions that leads to maximum average reward. 


The rest of the paper is organized as follows, In section~\ref{sysModel}, we describe our system model and problem formulation. In section~\ref{algo}, we explain our Reinforcement Learning based approach that we propose for solving the proposed formulation. In section~\ref{sim}, we describe in details our implementation and we also discuss the results. Finally, in section~\ref{conc}, we conclude the paper.

 \begin{figure}
\centering
\includegraphics[trim=1in 0.1in 0.3in 0in, clip, width=.48\textwidth]{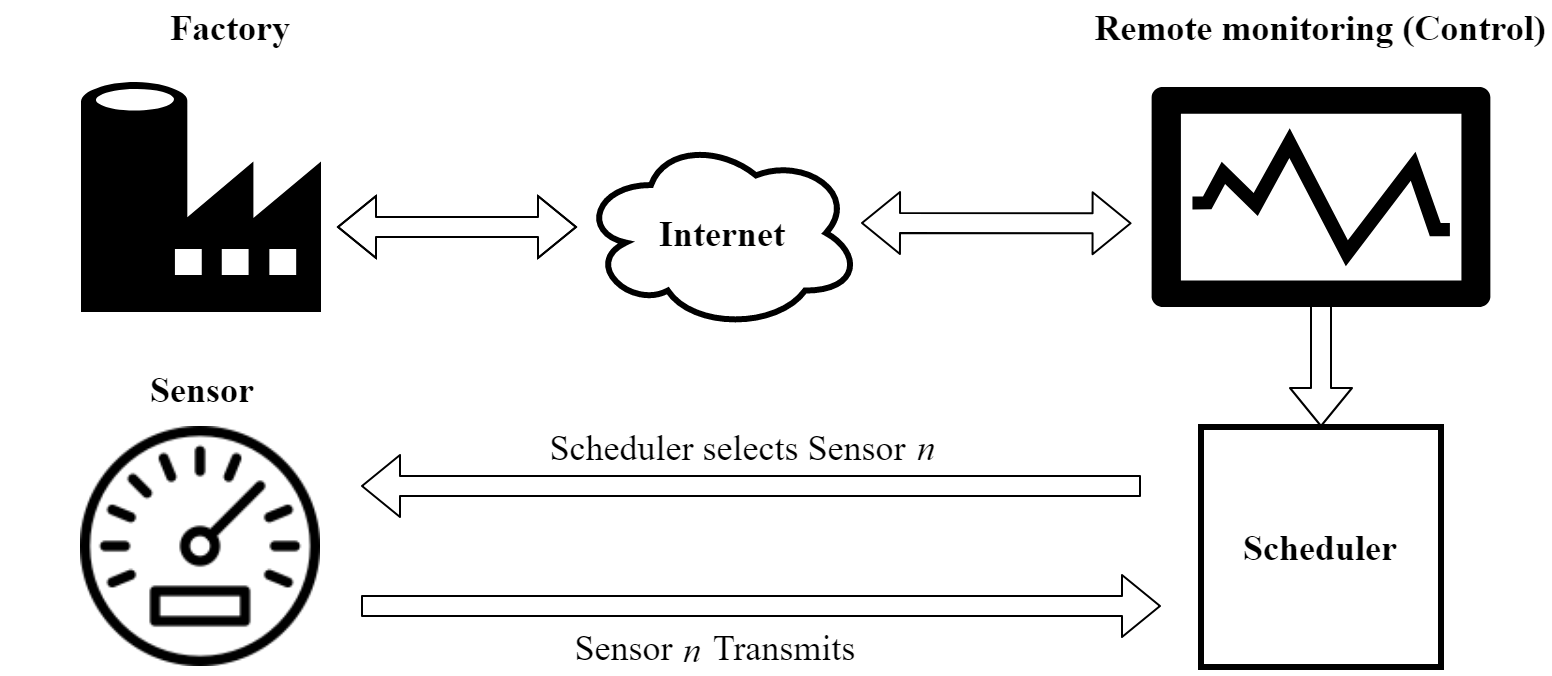}	
 \caption{System Model}
 \label{fig:sysModel}

\end{figure}

\section{System Model and Problem Formulation}
\label{sysModel}
The system model considered in this work focuses on a remote monitoring/automation scenario (Fig~\ref{fig:sysModel}) in which the monitor/controller resides in a remote site and allows one sensor at a time to update its state by generating fresh packet and send it to the monitor/controller. 
We assume that there are $N$ sensors, and only one sensor $n \in \{0,1, \cdots, N-1\}$ is chosen to update its state at job number $k \in \{0,1,\cdots, K-1\}$ by sending a packet of size $L_n$ Bytes at a rate $r(k)$ to the controller. Where $r(k)$ is the average rate at the time of job number $k$.

 Let $x^n(k)$ be an indicator variable that indicates the selection of sensor $n$ by the controller at job id $k$, i.e. $x^n(k)=1$ if the controller selects sensor $n$ at job id $k$, and $0$ otherwise. When  $x^n(k)=1$, sensor $n$ samples a new state, generates a packet accordingly, and sends it to the controller. The packet is successfully received by the controller with probability $p \in (0,1]$. Therefore, the packet is corrupted with probability $1-p$. Let $d_n(k)$ be the indicator function that is equal to the number of times sensor $n$ retransmits its packet at the job id $k$ before it is successfully received at the controller. For example, if the packet fails at the first transmission and then successfully received at the second transmission then $d_n(k) = 2$. Since, the controller selects only one sensor at any job id $k \in \{0,1,\cdots, K-1\}$, the following constraints must hold for $x^n(k), \forall n,k$:
 
  \begin{equation}
   \sum_{n=1}^{N} x^n(k)=1, \forall k \in \{0,1,\cdots, K-1\}
   \label{c1}
   \end{equation}
   \vspace{-5pt}
  \begin{equation}
   x^n(k) \in \{0,1\}, \forall n \in \{0,1,\cdots, N-1\},k \in \{0,1,\cdots, K-1\}
   \label{c7}
   \end{equation}
   
 
 Finally, let $a_n(k)$ be the age of the $n$-th sensor's information in seconds at the time of controller's job $k$. For example, if the $n$-th sensor has successfully transmitted a packet to the controller at the job $k-1$, the age of the $n$-th sensor's information at the controller will be $\sum_{i=1}^{d_n^k} L_n/r_i(k)$. Mathematically, the AoI of sensor $n$ at the time of job $k+1$, $a_n(k+1)$ evolves according to the following equation:
 
  \begin{align}
  &a_n(k+1)=x^n(k)\cdot \sum_{i=1}^{d_n^k}\frac{L_n}{r_i(k)} \nonumber\\&+ (1-x^n(k)) \cdot \Big(a_n(k)+\sum_{m=1, m\neq n}^{N}x^m(k)\cdot \sum_{i=1}^{d_m^k}\frac{L_m}{r_i(k)}\Big)
  \label{c3}
   \end{align}
Equation \eqref{c3} simply states that, the AoI for sensor $n$ at the time of job $k+1$ is equal to the time that is spent  transmitting a packet from sensor $n$ to the controller if sensor $n$ was selected in job $k$. Otherwise, if another sensor $m\neq n$ was selected in job $k$, the AoI of sensor $n$ at the controller side at time of job $k+1$ is equal to the AoI at the time of job $k$ plus the time that is spent transmitting a packet from sensor $m$ to the controller.
   
   To account for latency and reliability, we minimize an objective function that is a weighted sum of two terms. The first term is the sum average AoI, and the second term is the sum of the probability that AoI of each sensor $n$ at any job $k$ exceeds a predefined threshold. Therefore, the overall AoI minimization problem is:
   
   \begin{align}
\textbf{Minimize: }& \underset{K \rightarrow \infty}{\text {lim}}\frac{1}{KN}\sum_{n=0}^{N-1}\sum_{k=0}^{K-1}a_n(k)\nonumber\\&+\sum_{n=0}^{N-1}\lambda_n\sum_{k=0}^{K-1}{\textbf {Pr}}\{a_n(k) > \tau_n\}
 \label{mainObj}
 \end{align}
 
 $\quad\quad\quad\quad${\textbf{subject to}} \eqref{c1}-\eqref{c3}
 \vspace{0.1in}
 
where $\tau_n$ is the predefined threshold of sensor $n$.  In our objective function~\eqref{mainObj}, we assume that different sensors can tolerate different AoI thresholds. $\lambda_n$ is the weight of sensor n; higher the weight more is the penalty of exceeding the AoI's threshold. Finally, In order to optimize the tail performance, we should choose $\lambda_n \gg 1$, so that exceeding $\tau_n$ has a very high penalty and reduces the performance significantly.
   
{\bf Structure of the problem: } The proposed problem is a stochastic optimization problem with integer (non-convex) constraints. Integer-constrained problems even in the deterministic settings are known to be NP hard in general. Very limited problems in this class of discrete optimization are known to be solvable in polynomial time. Moreover, neither the packet drop probability nor the available bandwidth are known non-causally. Furthermore, all the sensors are competing over the available bandwidth. Therefore, we propose a learning based algorithm in order to learn and solve the proposed problem. In particular, we investigate the use of Reinforcement Learning (RL). In the next section, we describe our proposed algorithm for solving our formulated problem using RL.

 \section{Proposed Algorithm Based on Reinforcement Learning}
 \label{algo}
 In this paper, we consider a learning-based approach to find the scheduling policy from real observations. Our approach is based on reinforcement learning (RL). In RL, an agent interacts with an environment. At each job, the agent observes some state $s_k$ and performs an action $a_k$. After performing the action, the state of the environment transitions to $s_{k+1}$, and the agent receives a reward $R_k$. The objective of learning is to maximize the expected cumulative discounted reward defined as $E\{\sum_{k=0}^K\gamma^kR_k\}$. The discounted reward  is considered to ensure a long term system reward of the current action.
 
\begin{figure}
\centering
\includegraphics[trim=1in 0.4in 0.9in 0in, clip, width=.5\textwidth]{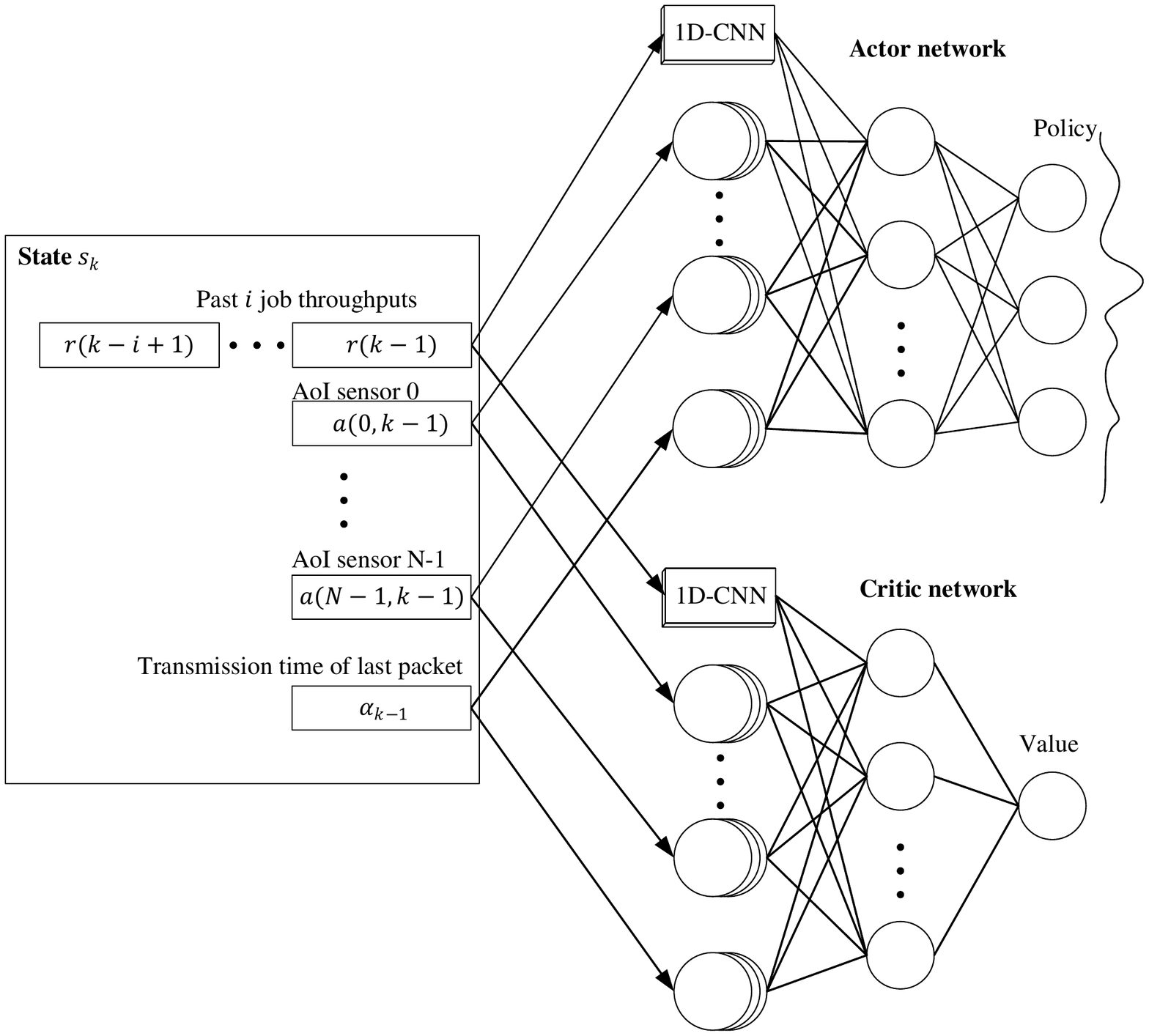}	
 \caption{The Actor-Critic Method that is used in our Proposed Scheduling Algorithm}
 \label{fig:act}

\end{figure}

 Our reinforcement learning approach is described in Fig.\ref{fig:act}. As shown, the scheduling policy is obtained from training a neural network. The agent observes a set of metrics including the current AoI of every sensor $a_n(k),\forall n,k$, and the throughput achieved in the last $j$ jobs and feeds these values to the neural network, which outputs the action. The action is defined by which sensor to choose for the next job. The reward is then observed and fed back to the agent. The agent uses the reward information to train and improve its neural network model. We explain the training algorithms later in the section. Our reward function, state, and action spaces are defined as follows:
 
 \begin{itemize}
 \item {\bf Reward Function}: The reward function at the end of job $k$, $R_k$ is defined by the following equation which is a sum of two terms. The first term is the sum of all sensors' AoIs at the end of job $k$, and the second term is a weighted sum of the penalties encountered when exceeding  AoI thresholds. 
 \begin{equation}
 R_k=-\sum_{n=0}^{N-1}a_n(k)-  \sum_{n=0}^{N-1}\lambda_n\cdot{\bf 1}_{(a_n(k) > \tau_n)}
 \end{equation}
 Where ${\bf 1}_{(\cdot)}$ is an indicator function. 
 
 \item {\bf State}: The state at job number $k$ is.
 \begin{itemize}
 \item The age of every sensor $n$ at job $k$, $a_n(k),\forall n,k$
 \item The throughput that was achieved for the last $j$ assigned jobs $r(k-j+1),\cdots,r(k-1)$.
 \item The time that was spent in downloading the last packet $\alpha_{k-1}$. which reflects the packet size as well as the current network conditions (packet drop and throughput).  
 \end{itemize}
  \item {\bf Action}: The action is represented by a probability vector of length $N$ such that if the $n$-th element is the maximum, sensor $n \in \{0,\cdots,N-1\}$ will be scheduled for job id $k$.
 \end{itemize}
 
 The first step toward generating the scheduling algorithm using RL is to run a training phase in which the learning agent explores the network environment. In order to train our RL based system, we use A3C \cite{mnih2016asynchronous} which is the state-of-art actor critic algorithm. A3C involves training two neural networks.
 
 Given state $s_k$ as described in Fig~\ref{fig:act}, the RL agent takes an action $a_k$ which corresponds to choosing one sensor for job $k$. The agent selects a certain action based on a policy which is defined as a probability over actions $\pi : \pi(s_k,a_k) \leftarrow [0,1]$. Specifically, $\pi(s_k,a_k)$ is the probability of choosing action $a_k$ given state $s_k$. The policy is a function of parameter $\theta$, which is referred to as policy parameter in reinforcement learning. Therefore, for each choice of $\theta$, we have a parametrized policy $\pi_\theta(s_k,a_k)$.
 
 After performing an action at job id $k$, a reward $R_k$ is observed by the RL agent. The reward reflects the performance of each action (sesnor selection) in the metric we need to optimize. Policy gradient method~\cite{sutton2000policy} is used to  train the actor-critic policy. Here, we describe the main steps of the algorithm. The main job of policy gradient methods is to estimate the gradient of the expected total reward. The gradient of the cumulative discounted reward with respect to the policy parameters $\theta$ is computed as: 
 
 \begin{equation}
 \nabla E_{\pi_{\theta}}\{\sum_{k=0}^{K-1}\sum_{t=0}^{K-1}\gamma^tR_{k+t}\}=E_{\pi_{\theta}}\{\nabla_\theta log \pi_\theta(s,a)A^{\pi_\theta}(s,a)\}
 \end{equation}
where $A^{\pi_\theta}(s,a)\}$ is the advantage function~\cite{mnih2016asynchronous}. $A^{\pi_\theta}(s,a)\}$ reflects how better an action compared to the average one chosen according to the policy. Since the exact $A^{\pi_\theta}(s_k,a_k)\}$ is not known, the agent samples a trajectory of scheduling decisions and uses the empirically computed advantage as an unbiased estimate of $A^{\pi_\theta}(s_k,a_k)\}$. The update of the actor network parameter $\theta$ follows the policy gradient which is defined as follows:
 
  \begin{equation}
\theta \leftarrow \theta + \alpha \sum_{k=0}^{K-1}\nabla_\theta log \pi_\theta(s_k,a_k)A^{\pi_\theta}(s_k,a_k)
\label{policyUpdate}
 \end{equation}
where $\alpha$ is the learning rate. In order to compute the advantage $A(s_k,a_k)$ for a given action, we need to estimate the value function of the current state ${\cal{V}^{\pi_\theta}}(s)$ which is the expected total reward starting at state $s$ and following the policy $\pi_\theta$. Estimating the value function from the empirically observed rewards is the task of the critic network. To train the critic network, we follow the standard temporal difference method~\cite{sutton2011reinforcement}. In particular, the update of $\theta_v$ follows the following equation:
 
   \begin{equation}
\theta_v \leftarrow \theta_v + \alpha^\prime \sum_{k=0}^{K-1}\big(R_k + \gamma V^{\pi_\theta}(s_{k+1},\theta_v)-V^{\pi_\theta}(s_k,\theta_v)\big)^2
\label{valueUpdate}
 \end{equation}
 
 Where $\alpha^\prime$ is the learning rate of the critic, and $V^{\pi_\theta}$ is the estimate of $\cal{V}^{\pi_\theta}$. Therefore, for a given ($s_k,a_k,R_k, s_{k+1}$), the advantage function, $A(s_k,a_k)$ is estimated as $R_k+\gamma V^{\pi_{\theta}}(s_{k+1},\theta_{v})-V^{\pi_{\theta}}(s_k,\theta_{v})$
 
 Finally, we would like to mention that the critic is only used at the training phase in order to help the actor converge to the optimal policy. The actor network is then used to make the scheduling decisions.

 \begin{figure*}
\centering
\includegraphics[trim=1in 0.4in 0.9in 0.6in, clip,  width=\textwidth]{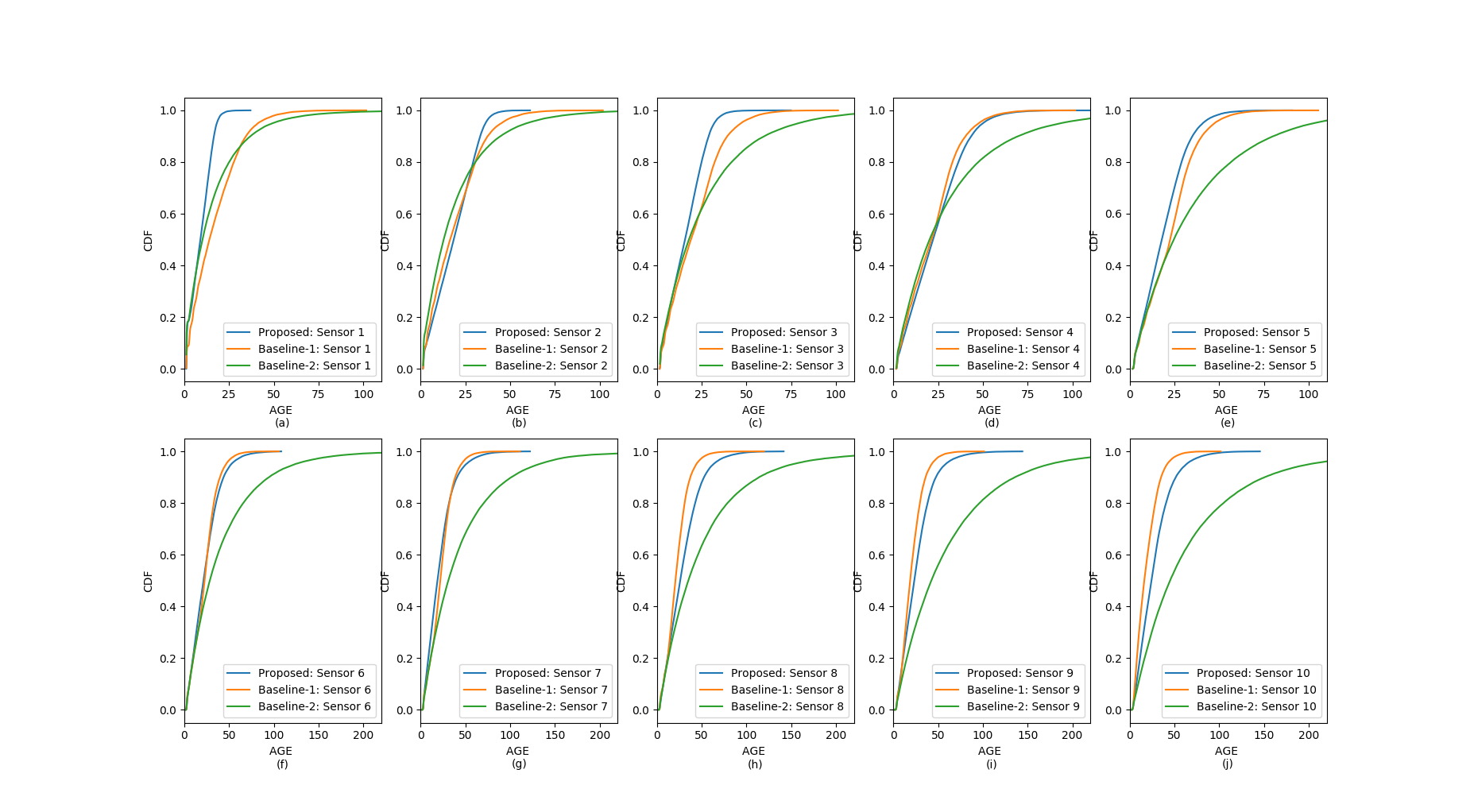}	
 \caption{CDF plots of the AoI of all sensors (a) sensor 0, $\cdots$, (j) sensor 9 }
 \label{fig:results}
\end{figure*}

\section{Simulation}
\label{sim}
\subsection{Implementation}
To generate our scheduling algorithm, we modified and trained the RL agent architecture described in \cite{mao2017neural} to serve our purpose.  The RL agent consists of an actor-critic pair. Both actor and critic network uses the same NN structure, except that the final output of the critic network is a linear neuron with no activation function. We pass the throughput that was achieved in the last 5 jobs to a 1D convolution layer (CNN) with 128 filters, each of size 4 with stride 1. The output of this layer is  aggregated with all other inputs in a hidden layer that uses 128 neurons with a relu activation function. At the end, the output layer (10 neurons) applies the softmax function. To account for discounted reward, we choose $\gamma = 0.9$. Moreover, we set the learning rates of both the actor and critic to $0.001$. We implemented this architecture in python using TensorFlow~\cite{abadi2016tensorflow}. Moreover, we used real bandwidth traces \cite{riiser2013commute,fccData,mao2017neural} to train our RL agent, and we set the packet drop probability to be $10\%$. One final thing, to ensure that the RL agent explores the action
space during training, we added an entropy regularization term to the actor?s update as described in \cite{mao2017neural}, and we initially set the entropy weight to $5$ in order to explore different policies. However, we kept generating models, pass them again to start over the training while reducing the entropy weight until we reached zero weight. We used the model that was generated with entropy weight being equal to zeros to generate the results.

We assume that there are $10$ sensors ($N=10$) generating packets of sizes, $50$ to $500$ bytes, with step size of $50$ bytes. The sensors have their AoI thresholds ranging from $30$ to $210$ms with a step size of $20$ms. Therefore, the first sensor (sensor 0) imposes a stricter AoI threshold, and the last one (sensor 9) has the much looser threshold value and higher packet size. We set $\lambda_n$ to be equal to $1000\cdot(N-n)/N$, where $n=0,\cdots, N-1$ represents the sensor ID. Therefore, we get more penalty when the AoI of the sensor that has the tighter threshold exceeds its target. 

For comparison, we considered two baselines, baseline 1 always chooses the sensor that has EDF to transmit which reflects the Earliest deadline First strategy. We refer to this baseline by ``{\it EDF}'' algorithm. In the other hand, baseline 2 is a modified version of Optimal Stationary Randomized Policy (OSRP) proposed in \cite{OptimAoI} that considers minimizing the probability of exceeding AoI thresholds. i.e, it randomly selects a sensor for job $k,\forall k$ with a probability of selection that is inversely proportional to its AoI threshold. Therefore, the sensor that has a stricter deadline is chosen more frequently. We refer to this baseline by ``{\it OSRP}'' algorithm

\subsection{Discussion}

Now we report our results using the test traces. throughout this section, we refer to our proposed algorithm by ``RL'' algorithm. The results are described in both table~\ref{compComplexity} and Fig~\ref{fig:results}. We clearly see from the total normalized objective function shown in table~\ref{compComplexity}, which is computed as per equation \eqref{mainObj} and normalized with respect to the ``RL'' algorithm, that our proposed RL based algorithm significantly outperforms baselines. The RL algorithm achieves the minimum objective value among the three algorithms. Its objective is around $50\%$, and $100\%$ lower than EDF (baseline 1) and OSRP (baseline 2) algorithms respectively. Moreover, RL algorithm achieves the minimum threshold violation for each sensor. For example, the AoI of the $0$-th sensor is not exceeding its threshold at any time using the proposed algorithm. However, the AoI of the same sensor exceeds its threshold $16\%$ of the time using baseline 1 and 2. Furthermore, the RL algorithm totally eliminates the violations for the sensors 4 onward. In the other hand, baseline 2 continues to experience considerably high threshold violation for all sensors.

We also notice that the RL algorithm maintains a trade-off between minimizing the average AoI of each user and minimizing the threshold violation which is very important requirement for URLLC. For example, for the first 3 sensors (sensors 0, 1, and 2), RL algorithm achieves the minimum AoI and minimum threshold violation. Note that the penalty of AoI violation is inversely proportional to the sensor's index. i.e, sensor 0's threshold violation degrades the performance of the algorithm much higher than sensor 1's violation and so on. We see for some sensors that EDF based approach (baseline 1) achieves a lower average AoI. However, that comes at the cost of having more threshold violation for sensors that have stricter AoI threshold and higher threshold violation penalty weight (e.g, sensor 0). Consequently, it degrades the performance more significant than the higher average AoI. Therefore, for the considered objective function which gives higher penalty to violating threshold AoI, our proposed RL based algorithm is able to learn that and to significantly outperform the other two algorithms.

Fig~\ref{fig:results}(a-j) plots the CDF of each sensor's AoI for the three algorithms. The CDF plots along side the results reported in table~\ref{compComplexity} show that RL algorithm outperforms the baselines in both average AoI and probability of exceeding the AoI threshold for the sensors with stricter deadlines Fig.~\ref{fig:results}(a-c). For example, in Fig.~\ref{fig:results}-(a), the CDF of the AoI of sensor 0 is plotted, which has the stricter threshold among all sensors. We see that RL algorithm achieves the minimum AoI for this sensor all the time compared to baseline 1 and 2.  The RL algorithm runs into higher average AoI than ``EDF'' algorithm (baseline 1) for sensors with looser AoI threshold, but that comes at the gain of achieving minimum threshold violation for most of the sensors. In conclusion, we clearly see that the RL based approach that is proposed in this paper learns how to consider minimizing the average AoI of every sensor while maintaining the probability of exceeding the AoI threshold of each sensor as low as possible. Moreover, it learns how to respect the weights specified by the objective function which gives much higher penalty to violating thresholds than achieving lower average AoI.  

\begin{table}[h!]
  \centering
  \caption{}
  \begin{tabularx}{0.48\textwidth}{|X|X|X|X|} \hline
     &RL& Baseline 1 (EDF) & Baseline 2 (OSRP) \\ \hline
  \textbf{Normalized Objective Eq\eqref{mainObj}}  &1&1.53  &2.1    \\ \hline
   \textbf{Pr AoI}$_0 >  \tau_0$ &0\%  &16.24\%  &16.18\%   \\ \hline
   \textbf{Pr AoI}$_1 >  \tau_1$ &0.06\%  &6.04\%  &7.80\%   \\ \hline
   \textbf{Pr AoI}$_2 >  \tau_2$ & 0.016\% & 2.46\% & 7.02 \%  \\ \hline
   \textbf{Pr AoI}$_3 >  \tau_3$ &0.06\%  & 0.04\% &5.55\%   \\ \hline
   \textbf{Pr AoI}$_4 >  \tau_4$ & 0\% &0\%  & 3.98\%  \\ \hline
   \textbf{Pr AoI}$_5 >  \tau_5$ &0\%  & 0\% & 4.32\%  \\ \hline
   \textbf{Pr AoI}$_6 >  \tau_6$ &0\%  & 0\% & 3.21\%  \\ \hline
   \textbf{Pr AoI}$_7 >  \tau_7$ &0\%  & 0\% & 3.63\%  \\ \hline
   \textbf{Pr AoI}$_8 >  \tau_8$ &0\%  &0\%  & 3.82 \%  \\ \hline
   \textbf{Pr AoI}$_9 >  \tau_9$ &0\%  &0\%  &4.42\%   \\ \hline 
   \textbf{Avg sensor 0} &9.82  &16.93  &15.89   \\ \hline
   \textbf{Avg sensor 1} &14.39  &19.12  &19.34   \\ \hline
   \textbf{Avg sensor 2} &17.30  &20.83  &26.31  \\ \hline
   \textbf{Avg sensor 3} &22.53  &22.05  &30.90   \\ \hline
   \textbf{Avg sensor 4} &20.07  &22.82  &35.35   \\ \hline
   \textbf{Avg sensor 5} &23.86  &23.12  &41.90   \\ \hline
   \textbf{Avg sensor 6} &22.90  &22.83  &45.09   \\ \hline
   \textbf{Avg sensor 7} &27.25  &22.08  &51.75   \\ \hline
   \textbf{Avg sensor 8} &24.28 &20.83  &61.27  \\ \hline
   \textbf{Avg sensor 9} &24.14  &19.09  &67.14   \\ \hline
   
  \end{tabularx}
  \label{compComplexity}
\end{table}
\section{Conclusion and Future Work}
\label{conc}
In this paper, we investigated the use of machine learning in solving network resource allocation/scheduling problems. In particular, we developed a reinforcement learning based algorithm to solve the problem of  AoI minimization for URLLC networks. We considered the system in which a number of sensor nodes are transmitting a time sensitive data to a remote monitoring side. We optimize a metric that maintains a trade-off between minimizing the sum of the expected AoI of all sensors and minimizing the probability of exceeding a certain AoI threshold for each sensor. We trained our reinforcement learning algorithm using the state-of-the-art actor-critic algorithm over a set of public bandwidth traces with non zero probability of packet drop. Simulation results show that the proposed algorithm outperforms the considered baselines in terms of optimizing the considered metric. Investigating the switching between sensors at time slot level to minimize the Maximum Allowable Transfer Interval (MATI) for control over wireless is an interesting direction for future work.

\bibliographystyle{IEEEtran} %
\bibliography{ref}

\end{document}